\documentstyle[12pt,epsfig]{article}
\input{epsf}
\begin{document}
\setlength{\baselineskip}{0.30in}
\newcommand{\be}{\begin{eqnarray}}
\newcommand{\ee}{\end{eqnarray}}
\newcommand{\bi}{\bibitem}
\newcommand{\nue}{\nu_e}
\newcommand{\num}{\nu_\mu}
\newcommand{\nut}{\nu_\tau}
\newcommand{\nus}{\nu_s}
\newcommand{\mne}{m_{\nu_e}}
\newcommand{\mnm}{m_{\nu_\mu}}
\newcommand{\mnt}{m_{\nu_\tau}}
\newcommand{\lar}{\leftarrow}
\newcommand{\rar}{\rightarrow}
\newcommand{\lrar}{\leftrightarrow}
\newcommand{\nuh}{\nu_h}
\newcommand{\mnh}{m_{\nu_h}}
\newcommand{\taut}{\tau_{\nut}}
\newcommand{\fg}{f_{\gamma}}
\newcommand{\dt}{\partial_t}
\newcommand{\pr}{\partial}
\newcommand{\rotv}{{\bf \nabla}\times {\bf v}}

\begin{center}
\vglue .06in
{\Large \bf  
Generation of large scale magnetic fields at recombination epoch  }
\bigskip
\\
{\bf Z. Berezhiani}$^{a}$ and 
{\bf A.D. Dolgov}$^{b,c}$ \\
$^a$ {\it Dipartimento di Fisica, Universit\`a di L'Aquila, 67010 Coppito,
L'Aquila, and \\
INFN, Laboratori Nazionali del Gran Sasso, 67010 Assergi, L'Aquila, 
Italy;
e-mail: berezhia@fe.infn.it
 } \\
$^b$ {\it{INFN, sezione di Ferrara, Via Paradiso, 12 - 44100 Ferrara,
Italy} \\
$^c$ {\it ITEP, Bol. Cheremushkinskaya 25, Moscow 113259, Russia;
e-mail: dolgov@fe.infn.it
}}\\
\end{center}

\vspace{.3in}
\begin{abstract}

It is argued that large scale cosmic magnetic field could be generated 
in the primeval plasma slightly before hydrogen recombination. Non-zero 
vorticity, necessary for that, might be created by the photon diffusion 
in the second order in the temperature fluctuations. The resulting seed 
fields at galactic scale would be only 4 orders of magnitude smaller than 
the observed ones and with a mild galactic dynamo amplifying the seed 
fields by the factor $\sim 10^4$ an existence of coherent magnetic fields
in galaxies may be explained.

\end{abstract}


\section{Introduction \label{s-intr}}

Existence of magnetic fields on astronomically large scales remains one of 
unsolved cosmological mysteries. It is known from observations that there
are magnetic fields in galaxies with the strength about micro-gauss which
are homogeneous over galactic size $l_{gal} \sim$ ( a few) kpc,
and maybe, even more puzzling, 
intergalactic magnetic fields with the strength approximately three orders of
magnitude weaker but homogeneous on much larger intergalactic scale of
hundreds kpc (for review see e.g. refs.~\cite{magn-rev}). Though in stellar 
processes much stronger magnetic fields can be generated, their coherence 
scale is negligible in comparison with $l_{gal}$. There are two competing
ideas of explanation of the phenomenon~\cite{magn-rev,grasso-rub01}. 
The first one is based on traditional astrophysics and pursues the possibility
that strong magnetic fields generated in multiple stellar catastrophes could 
be ejected into interstellar space and by field line reconnection might
create magnetic fields coherent on galactic size (for a recent review
see ref.~\cite{biermann}). However there are no compelling
quantitative arguments in favor of this hypothesis. Moreover, the energy
density of galactic magnetic fields are of the same magnitude as the energy
density of cosmic microwave background radiation (CMBR) so its magnitude is 
about $10^{-10}$ of the total galactic mass/energy. Such a huge contribution 
is difficult to explain by conventional mechanisms. It is even more difficult
to explain in this way intergalactic magnetic fields, if they exist.

Other possible mechanisms of generation of galactic magnetic fields are
based on physical processes in the early universe - for the review
see e.g. refs. \cite{grasso-rub01,dolgov01}. One can also find an
extensive list of literature in ref. \cite{dimopoulos02}.
Basically there are three different mechanisms discussed: \\
1. Breaking of conformal invariance of electromagnetic interaction at 
inflationary stage. The latter could be realized either through new
non-minimal (and possibly non gauge invariant) coupling of 
electromagnetic field to curvature~\cite{turner88}, or in dilaton
electrodynamics~\cite{ratra92}, or by conformal anomaly 
in the trace of the stress tensor induced by quantum corrections
to Maxwell electrodynamics~\cite{dolgov93}.
\\
2.  First order phase transitions in the early universe~\cite{hogan83}
producing bubbles of new phase inside the old one. A different mechanism
but also related to phase transitions is connected with topological
defects, in particular, cosmic strings~\cite{vachaspati91}. A recent
discussion and a model of generation of large scale magnetic field can 
be found in ref.~\cite{boyan02}.
\\
3.  Creation of stochastic inhomogeneities in cosmological charge 
asymmetry, either electric~\cite{dolgov-s93}, or e.g. 
leptonic~\cite{dolgov-gr01} at 
large scales which produce turbulent electric currents and, in turn, 
magnetic fields.

In this work we will consider a new mechanism somewhat similar to those
mentioned in point 3 above but in contrast to them this mechanism does
not demand any new physics and could be realized at relatively late stage
of the universe evolution, namely, at red-shifts $z\sim 10^3-10^4$, near
hydrogen recombination. The mechanism suggested here is intermediate
in time between the early universe ones and the astrophysical mechanism
which took place practically in contemporary universe after galaxies were
formed and stellar explosions took place.

The basic features of the suggested model of magnetic field generation
are the following. We show that, despite low Reynolds number (\ref{R}),
non-zero, though small, vorticity can be generated in the cosmic
electron-photon fluid due to inhomogeneity of the latter and especially
due to different spectrum of inhomogeneities of electronic and photonic
components. Such difference could be created at relatively late stage 
of cosmological evolution (near hydrogen recombination) even from initially 
adiabatic density perturbations. The motion of cosmic plasma would
create electric currents because of different velocities of electrons and
protons and since the vorticity of the motion would be non-vanishing the
same would be true for the currents, 
${\bf \nabla}\times {\bf J} \neq 0$. 
Such currents are known to generate magnetic fields.
An attractive feature of the proposed mechanism is that no new physics
has to be invoked. The mechanism operates in the standard cosmological
model in the frameworks of the usual Maxwell electrodynamics. In what follows
we will estimate the magnitude of the generated magnetic field and show
that it can be strong enough and have sufficiently large coherence scale
so that after dynamo amplification (for the review of the latter see
refs.~\cite{dynamo}) it can
explain the observed fields in the galaxies and, possibly, even 
intergalactic ones.

\section{Hydrodynamics of cosmic plasma at recombination \label{s-hydro}}

Fluid motion under pressure forces is governed by the hydrodynamical
equation (see e.g. the book~\cite{ll-hydro}):
\be
\rho \left(\dt v_i + v_k \pr_k v_i \right) =
-\pr_i p + \pr_k \left[\eta\,\left(\pr_k v_i + \pr_i v_k - 
{2\over 3} \delta_{ik} \pr_j v_j \right) + \pr_i \left( \zeta\, \pr_j v_j
\right) \right] 
\label{master-eq}
\ee
where $v$ is the velocity of the fluid element, $\rho$ and $p$
respectively are the energy and pressure densities of the fluid, 
and $\eta$ and $\zeta$ are the first and second viscosity coefficients.
In the case of constant viscosity coefficients this equation is reduced to the
well known Navier-Stokes equation. The coefficient $\eta$ is related to the
mean free path of particles in fluid as 
\be
\eta /\rho \equiv \nu = l_{f}
\label{lmfp}
\ee 
In what follows we disregard second viscosity $\zeta$.

The character of the solution to eq.~(\ref{master-eq}) crucially 
depends upon the value of the Reynolds number
\be
R_\lambda = \frac{v \lambda}{\nu} 
\label{R}
\ee
where $\lambda$ is the wavelength of the velocity perturbation.
If $R\gg 1$, then the fluid motion would become turbulent and non-zero 
vorticity would be created by spontaneously generated turbulent eddies. 
In the opposite case of low $R$ the motion is smooth and in the case
of incompressible fluid with homogeneous $\rho$ and viscosity coefficients
$\eta$ and $\zeta$ the fluid velocity would have potential character with
vanishing vorticity, ${\bf\nabla} \times {\bf v} =0$. 
However, the above assumptions of homogeneity is not precise 
and some vorticity can be generated even with low $R$.

Let us consider now the cosmological epoch before the hydrogen 
recombination, when the plasma consists of three components:
photons, electrons, and baryons mainly constituted by protons 
(for simplicity, we shall neglect small amount of $^4$He nuclei).  
For temperatures $T\geq 1$ eV the energy density of 
$(e,p,\gamma)$-plasma is dominated by photons. 
Indeed, at the present day the CMBR energy density is 
$\rho_{\gamma} \approx 0.26$ eV/cm$^{3}$,
i.e. approximately $10^3$ smaller than the energy density of 
baryonic matter $\rho_b$, and thus 
$\rho_b$ and $\rho_{\gamma}$ become comparable at red-shift 
$z\approx 10^3$ or at $T\approx 0.23$ eV.\footnote{ Throughout 
this paper, we use the following values of the cosmological 
parameters: the CMBR temperature $T^{(0)}_{\gamma} = 2.725$ K, 
the baryon-to-photon ratio $\beta= n_B/n_\gamma = 6\times 10^{-10}$, 
the Hubble constant $h = 0.71$,  
the matter density of the universe $\Omega_m h^2 = 0.135$, 
and the baryon density $\Omega_b h^2 = 0.0224$. 
  }    
The contribution of dark matter (DM) into energy density is not 
important for plasma hydrodynamics as far as the DM does not interact 
with plasma.
For the moment, we also neglect cosmological expansion considering 
characteristic times smaller than the Hubble time $H^{-1}$ 
at the appropriate epoch:  
\be  
H^{-1}  = \frac {27 ~ {\rm kpc}} 
{ T_{\rm eV}^{3/2}\, [T_{\rm eV} + 0.76]^{1/2} } \, ,     
\label{H} 
\ee
where $T_{\rm eV} = (T/1~ {\rm eV})$. At the present scale, 
it corresponds to 
$(1+z) H^{-1} = 110\, T_{\rm eV}^{-1/2} [T_{\rm eV} + 0.76]^{-1/2}$ Mpc.
This result is obtained for cosmological relativistic matter consisting
from photons and neutrinos (contributing 68\% with respect to photons); 
the cosmological constant is not essential.  
For labeling the cosmological epoch we will interchangeably
use the temperature $T$ or red-shift $1+z = 4260 \, T_{\rm eV}$.  

As far as the cosmological plasma was dominated by photons, 
the viscosity coefficient $\nu$ is determined by the photon 
mean free path:
\be
\nu = l_{\gamma} = \frac{1}{\sigma_T n_e X_e } \approx
\frac{ 30 ~ {\rm pc} } {X_e(T) T_{\rm eV}^{3} } \, ,  
\label{nu}
\ee
where $\sigma_T = 8\pi \alpha^2 /3 m_e^2 = 6.65\times 10^{-25}$ cm$^2$ 
is the Thomson cross-section, $n_e = \beta n_\gamma = 6\cdot 10^{-10}
\cdot 0.24 T^3$ is the electron number density, 
and $X_e$ is a fraction of the free electrons: $X_e(z)$ is practically 
1 for $z > 1500$, and sharply decreases for smaller $z$'s, reaching 
values $\sim 10^{-5}$ at $z<1000$.
For $T \geq 1$ eV the mean free path (\ref{nu}) is much smaller than
the Hubble horizon:
\be
\frac{l_{\gamma}}{H^{-1}} \approx 
\frac{1.1 \times 10^{-3} } {X_e(T) T_{\rm eV} }\, 
\left[1+ 0.76\,T_{\rm eV}^{-1}\right]^{1/2}   
\label{l/t}
\ee
This ratio becomes comparable to unity only at temperatures 
$T \sim 0.3$ eV, when $X_e \ll 1$. 
At the present day scale the photon mean free path (\ref{nu})
would be 
$l_{\gamma}^{(0)} \approx 130\,\, {\rm kpc}\, X_e^{-1} T_{\rm eV}^{-2}$.    

Let us first estimate the Reynolds number of the fluid motion created 
by the pressure gradient in eq. (\ref{master-eq}). 
To this end we assume that the liquid is quasi incompressible and
homogeneous, so that the second term in the r.h.s. of this equation 
can be neglected. This is approximately correct and the obtained 
magnitude of fluid velocity is sufficiently accurate. 
In this approximation eq. (\ref{master-eq}) reduces to a much 
simpler one:
\be
\dt {\bf v} + \left( {\bf v}\, {\bf \nabla} \right) {\bf v} -
\nu \Delta {\bf v} = -{{\bf \nabla} p \over \rho } 
\label{ns-eq}
\ee

A comment worth making at this stage. The complete system of equations
includes also continuity equation which connects the time variation of 
energy density with the hydrodynamical flux (see below eq.~(\ref{contin}) 
and the Poisson equation for gravitational potential induced by density 
inhomogeneities. We will however neglect the gravitational force and the
back reaction of the fluid motion on the density perturbation. This
approximation would give a reasonable estimate of the fluid velocity
for the time intervals when acoustic oscillations are not yet developed,
i.e. for $t<\lambda/v_s$, where $\lambda$ is the wave length of the 
perturbation and $v_s$ is the speed of sound (in the case under 
consideration $v_s^2 = 1/3$). In fact the wave length should be larger
than the photon mean free path, to avoid diffusion damping, and 
the characteristic time interval, as we see below, should be  
somewhat larger than $\lambda$. So we may hope that our estimates of the
velocity are reasonable enough. Neglecting gravitational forces, especially
those induced by dark matter would result in a smaller magnitude of the
fluid velocity, so the real effect should be somewhat larger.

For small velocities (or sufficiently small wavelengths) we may neglect
the second term in the l.h.s. with respect to the third one. 
In this approximation the equation becomes linear and can be easily solved 
for the Fourier transformed quantities. 
Assuming that the parameters are time-independent (though it
is not necessary) we obtain: 
\be
{\bf v_k} =- {i{\bf k} \over 3k^2 \nu}\,\delta_k\,
\left[ 1 - \exp (- \nu k^2 t) \right] 
\label{v0}
\ee
where $\delta_k = (\delta \rho/\rho)_k$ is the Fourier transform of 
relative density perturbations, $\delta \rho /\rho$; 
its natural value is $\sim 10^{-4}$, though it might be much larger 
at small scales. The coefficient $1/3$ comes from
equation of state of relativistic fluid, $p=\rho/3$.

Therefore, for the Reynolds number we obtain:
\be
R_k = \frac{\delta_k}{3 \left( k\,\nu \right)^2 } 
\left[1 -\exp \left( - {\nu k^2 t } \right) \right]  \, . 
\label{R1}
\ee
If $\delta_k$ is weakly dependent on $k$, then $R_k$
is a monotonically rising function of the wavelength $\lambda = 2\pi/k$.  
For $t \ll \lambda^2/\nu $ it takes the value  
\be
R_k  = { t \over 3\nu} \, \delta_k
\label{R-max}
\ee 

Therefore, considering the perturbation 
with the wavelength $\lambda = 2\pi/k$ which enters the horizon 
at the time $t \sim H^{-1}$, we find that the  
maximum value of the Reynolds number is given by 
$R^{\rm max}_k \approx (H^{-1}/3l_\gamma) \delta_k $. 
For later times, the Reynolds number for the comoving wavelength 
decreases approximately as $\propto T$ until the density perturbation 
is completely damped by the photon diffusion. 
Thus, for $T \gg 1$ eV, using eq. (\ref{l/t}), we obtain  
$R^{\rm max}_k \sim 300 \delta_k \,T_{\rm eV} $, 
and so for the development 
of turbulence one needs $\delta_k\, T_{\rm eV} > 0.1$. 
This condition can be satisfied either at high temperatures, 
$T \geq 100$ eV, or for large density perturbations, 
$\delta_k \gg 10^{-4}$. This is not the case
for the adiabatic fluctuations with a nearly flat spectrum, 
as is generically predicted in inflationary scenarios.  
However, one cannot exclude a situation that at smaller 
wavelengths, corresponding to the present day scales $\leq 100$ kpc, 
there are substantial isocurvature fluctuations, 
e.g. the baryon density perturbations $(\delta \rho/\rho)_b$ 
are larger than  
the dark matter fluctuations $(\delta \rho/\rho)_{\rm DM}$.   

The baryon density perturbation of the wavelength $\lambda$,  
after many oscillations,  will be damped by the Silk effect~\cite{silk68}  
at the effective timescale $t_{\rm eff} \sim \lambda^2/\nu$. 
However, because of a large Reynolds number, it could happen
that during this time the photon diffusion dragging 
electrons would also produce non-zero vorticity.\footnote{  
The case of large Reynolds numbers, that might be created at earlier 
epoch, $T \sim 1 $ MeV, due to large leptonic asymmetries,
was considered in ref.~\cite{dolgov-gr01}. }
In this case the generation of magnetic fields might be strongly 
amplified. 
We shall keep in mind an interesting possibility of large 
$\delta_k$ at smaller scales, 
however now we shall mostly concentrate 
on a more plausible possibility of small Reynolds numbers 
and non-turbulent flow of cosmic fluid in  
$T\sim 1$ eV range of temperatures.
  
At first sight vorticity in laminar fluid motion is not generated in
the approximation given by eq. (\ref{ns-eq}). 
However, this is not quite so. 
If both $p$ and $\rho$ are different functions of space points, 
pressure gradient may create  motion with non-vanishing $\rotv$. 
Indeed, from eq. (\ref{ns-eq}) follows
\be
\dt {\bf \Omega} - \nu \Delta\, {\bf \Omega} = 
- {\bf \nabla} \times \left({ {\bf\nabla} p \over \rho }\right)
\label{dt-omega}
\ee
where ${\bf \Omega} = \rotv$ and we assume that velocity is small 
so that the term quadratic in $v$ was neglected. 
If the r.h.s. is non-vanishing, then
${\bf \Omega}$ would be non-zero too. However usually pressure density is 
proportional to the energy density, $p = w\rho$, with a constant
coefficient $w$ and hence
${\bf \nabla} \times \left({ {\bf \nabla} p / \rho }\right) =0$.

In the case under consideration, electrons are strongly coupled 
to photons.  Their effective mean free path is 
\be
l_{e} =  {\sqrt{3m_e/T} \over  \sigma_{\rm Th} n_\gamma}
\label{le}
\ee
The factor in the numerator takes into account large thermal momentum
of electrons, $\langle p \rangle \sim \sqrt{3m_eT}$, so that they need 
many collisions, with momentum transfer $\Delta p = T$, 
for a significant change of their momentum. Still despite
the presence of this factor, the 
mean free path of electrons is much smaller than 
that of photons simply because there are more than billion photons 
per a single electron in the plasma. 
So electrons are practically frozen in the plasma,
while photons may diffuse to much larger distance. 
One should keep in mind that the plasma must be locally electrically
neutral and thus redistribution of electron inhomogeneities is 
accompanied by the same redistribution of baryonic ones. 
This makes the mean free path of charged species of matter even smaller.
On the other hand, a motion of homogeneous component of electron number
density does not require dragging of baryons.

For the wavelengths larger than mean free path of charged particles
we may consider plasma as a fluid where charged particles are 
strongly coupled to photons. 
So for the estimate of the rotational velocity we will use the hydrodynamic
equation (\ref{dt-omega}). As we have already mentioned, vorticity, 
$\rotv$ could be non-zero only if the fluid is non-homogeneous and strictly 
speaking we should use eq. (\ref{master-eq}) with all parameters depending
upon space points. However, we believe that for a simple estimate 
eq. (\ref{dt-omega}) may be sufficient. Since before recombination the
interaction rate between radiation and charged particles is very high,
plasma/liquid should be in local thermal equilibrium with all constituents
having the same temperature $T(x)$. If $T$ would be the only parameter
which determines the state of the medium, then vorticity would 
not be generated
because we would have in our disposal only ${\bf\nabla} T$ and 
it is impossible to construct non-vanishing $\rotv$ from the
gradient of only one scalar function.
However, distributions of charged particles depend upon one more function,
their chemical potential: 
\be
f = \exp \left[ -{E\over T(x)} + \xi (x) \right]
\label{f}
\ee
where the dimensionless chemical potential 
$\xi$ can be readily expressed through
particle number density $n_e \approx n_B = \beta (x) n_\gamma$ with 
$\beta (x) = 6\cdot 10^{-10} + \delta \beta (x)$:
\be
\xi (x) = \ln \beta (x) + {\rm const} 
\label{xi-x}
\ee

Hence we will find that the source term in the vorticity 
equation (\ref{dt-omega}) is equal to:
\be
S_k\equiv -\epsilon_{ijk}\, \pr_j \left({\pr_i p \over \rho}\right) =
\epsilon_{ijk}\, { \pr_i \rho_\gamma \over 3 \rho_{\rm tot}}\,
{\pr_j \beta \over \beta}\, {\rho_b \over \rho_{\rm tot}} 
\label{source}
\ee
An essential feature here is that the spatial distribution of charged 
particles does not repeat the distribution of photons and hence the
vectors ${\bf \nabla} \rho_\gamma$ and  ${\bf \nabla} \beta$ are not 
collinear. 
This could occur if for lambda corresponding to subgalactic scales,
there exist baryon isocurvature fluctuations and thus
$\rho(x)$ and $\beta(x)$ have different profiles.
As we have mentioned above, different mean free paths of
photons and charged particles would maintain such non-collinearity of
the order of unity at the scales $\lambda \sim l_{\gamma} $.  
Moreover, even in the case of adiabatic perturbations
a shift in the distribution of photons and charged particles 
could also be created 
because of acoustic oscillations that proceeded with different phases of
radiation and matter densities. At the scales 
$\lambda \leq l_{\gamma}$
perturbations in the the plasma temperature would be erased by the
diffusion damping~\cite{silk68}, while for $\lambda \gg l_{\gamma}$
diffusion processes are not efficient and one would expect self-similar
perturbation leading to collinearity of 
${\bf \nabla} \rho_\gamma$ and  ${\bf \nabla} \beta$. On the other hand,
when $\lambda$ entered under horizon acoustic oscillations begun which
destroyed the self-similarity.
Thus the expected wavelengths of vorticity perturbations should be
between $l_{\gamma} <\lambda < H^{-1}$.

We assume that there is no additional suppression of the source 
term (\ref{source}) and by the order of magnitude its amplitude 
corresponding to the wavelength $\lambda$ can be evaluated as 
\be
S\sim {16\pi^2\over 3\lambda^2}\,\left({\delta T \over T}\right)_\lambda 
\, \left({\delta \beta \over \beta}\right)_\lambda\, 
\left({\rho_b \over \rho_{tot}}\right) \sim
10 \left({\delta T \over T}\right)_\lambda\, 
\left({\delta \beta \over \beta}\right)_\lambda\, T_{\rm eV}^{-1}  
\label{S}
\ee
Taking $\delta T/T \sim \delta \beta /\beta \sim 3\cdot 10^{-5}$ 
we obtain $ S\sim 10^{-8} T_{\rm eV}^{-1}$.

Now equation (\ref{dt-omega}) can be solved in the same way as 
eq.~(\ref{ns-eq}) and we find:
\be
|{\bf\Omega}| = {0.27\over l_{\gamma}\, T_{\rm eV} }\, 
\left({\delta T \over T}\right)_\lambda\, 
\left({\delta \beta \over \beta}\right)_\lambda\, 
\left[ 1 - 
\exp \left( - {4\pi^2\,l_{\gamma} t \over \lambda ^2}\right)\right]
\label{omega}
\ee
with $l_{\gamma}$ given by eq.~(\ref{nu}).

Vorticity could also be generated even if perturbations in plasma are
determined by a single scalar function, for example, by $T(t,{\bf x})$
because if might be proportional to the product 
$\partial_i T(t,{\bf x})\,\partial_j T(t',{\bf x}) $. These two gradients
generally are not collinear if taken at different time moments $t$ and
$t'$. To see that, let us start from the Boltzmann equation for
the distribution function $f(t,{\bf x},E,{\bf p})$ of photons:
\begin{equation}
\left( \frac{\partial}{\partial t} + {\bf V}\cdot{\bf \nabla} -
H\, {\bf p} \frac{\partial}{\partial {\bf p}} +
{\bf F}\, \frac{\partial}{\partial {\bf p}  } \right) 
f(t,{\bf x},E,{\bf p})
= I_{\rm coll}\left[f_{a}\right] ~, 
\label{boltz}
\end{equation}
where ${\bf V} = {\bf p}/E$ is the particle velocity (not to be confused
with the velocity ${\bf v}$ of macroscopic motion of the medium),
for photons $V=1$, while $v \ll 1$,
$E$ and ${\bf p}$ are respectively the particle energy and
spatial momentum, $H$ is the universe expansion rate, ${\bf F}$ is
an external force acting on particles in question (the latter is assumed
to be absent), and $I\left[f_{a}, f_b,...\right]$ is the collision 
integral depending on the distributions $f_a$ of all participating 
particles (for the definition of the $I_{\rm coll}$
see e.g. eq.~(43) of review~\cite{dolgov02}).

At temperatures in eV-range only the Thomson scattering of photons on 
electrons is essential, so the collision integral is dominated by the
elastic term. Integrating both parts of eq.~(\ref{boltz}) over 
$d^3 p/(2\pi)^3$ we arrive to the continuity equation:
\be
\dot n ({\bf x}) + {\bf \nabla}{\bf J} = 0
\label{contin}
\ee
where ${\bf J}$ is the photon flux given by
\be
{\bf J } \equiv {\bf v} n = \int {d^3 p \over (2\pi)^3}\,
{{\bf p}\over E}\, f
\label{J}
\ee
and ${\bf v}$ is the average macroscopic velocity of the photon plasma.
Using the standard arguments one can derive from eq.~(\ref{contin}) the
diffusion equation:
\be
\dot n = D\, \Delta n
\label{difus}
\ee
where $D\approx l_{\gamma}/3$ is the diffusion coefficient.
We will use this equation below to determine time evolution of the photon
temperature $T$.

If the elastic reaction rate
$\Gamma_{\rm el} = \sigma_{\rm Th} n_e X_e = 1/l_{\gamma}$ 
is sufficiently large, local thermal equilibrium would be established
and the photon distribution would be approximately given by
\be
f \approx f_{0} = 
\exp \left(- E/T +\xi \right)
\label{f0}
\ee
where the temperature and effective chemical potential could be
functions of time and space coordinates: $T=T(t,{\bf x})$ and
$\xi=\xi(t,{\bf x})$, and the photon mean free path is given by
eqs.~(\ref{nu},\ref{l/t}). Evidently $f_0$ annihilates the collision 
integral. We can find correction to this distribution, $f =f_0 + f_1$,
substituting this expression into kinetic equation (\ref{boltz}) and 
approximating the collision integral in the usual way as 
$-\Gamma_{\rm el} f_1$:
\be
\left( K  + \Gamma_{\rm el}\right) f_1 = - K f_0
\label{Df1}
\ee
where $K$ is the differential operator, 
$K = \partial_t + \left({\bf V} \,{\bf \nabla}\right)$.
The solution of this equation is straightforward:
\be
f_1 \left(t,{\bf x},E,{\bf V}\right) = - \int_0^t d\tau_1 
\exp{ \left[ -\int_0^{\tau_1} d\tau_2 
\Gamma_{el} \left(t-\tau_2, {\bf x}-{\bf V}\tau_2 \right)\right]}
K f_0\left(t-\tau_1, {\bf x} -{\bf V}\tau_1
\right)
\label{f1}
\ee
Using this result we can calculate the average macroscopic
velocity of the plasma. The calculations are especially simple if
elastic scattering rate is high and the integrals are dominated by
small values of $\tau_1$. In this case we obtain: 
\be
{ v_j}(t,{\bf x}) = {\int d^3 p { V_j} f_1(t,{\bf x},E,{\bf V})
\over \int d^3 p f_0(t,{\bf x},E) }
\label{v}
\ee
and its vorticity, $\Omega_i=\epsilon_{ijk} \partial_j v_k$:
\be
\Omega_i \approx 6\epsilon_{ijl}\, l_{\gamma}^2\, 
\left({\partial_j T \over T}\right)\, 
\left({\partial_l \partial_t T \over T}\right)
\label{omega-i}
\ee

To estimate time derivatives of the temperature we will use the diffusion 
equation~(\ref{difus}), from which we find $\partial_t T = D \Delta T$
and finally obtain for vorticity at the scale $\lambda$:
\be
|{\bf \Omega}|_\lambda 
\approx 2\, \left({\delta T \over T}\right)^2_\lambda\,
l_{\gamma}^3\,k^4 \approx 3\cdot 10^3
\left({\delta T \over T}\right)^2_\lambda\
{l_{\gamma}^3\over \lambda^4}
\label{omega-fin}
\ee
Since the photon diffusion erases temperature fluctuations at the scales
$\lambda < l_\gamma$ the vorticity has the maximum near 
$\lambda \sim l_\gamma$. This magnitude of vorticity is considerably
larger than found previously (\ref{omega}) and we will rely on it in
the estimates of magnetic field presented in the following section.

To avoid confusion let us mention that at the moment when perturbation
with a given wave length enters the horizon, $\lambda \gg l_\gamma$.
Later $(\lambda/l_\gamma)$ scales as $\sim a(t)^{-2}$ and till $\lambda$ 
remains larger than $l_\gamma$ the amplitude of perturbations does not
decrease significantly and only when $\lambda \leq l_\gamma$ the
photon diffusion damps density perturbations.

\section{Electrodynamics of cosmic plasma with helical flows and
magnetic field generation \label{s-elec}}

Generation and evolution of magnetic field strongly depends upon the 
electric conductivity of plasma which can be estimated as follows. 
Equation of motion of a charged particle in external electric field 
${\bf E}$ has the usual form
\be
m \dot{\bf V}_d  = e {\bf E}
\label{m-dotv}
\ee
So the drift velocity gained during the time $\Delta t$ is equal to
$V_d = e E \Delta t /m$. The charged particles (electrons)
keep on to be accelerated approximately during time between collisions, 
$\Delta t = l_{e}/ V_T$ where $V_T = \sqrt {3T/m}$ is the thermal 
velocity and the electron mean free path $ l_{e}$ is given by
eq.~(\ref{le}). After collision the particle loses (forgets) its
previous velocity and the process repeats. 
This is true for sufficiently weak fields when the drift velocity 
is small in comparison with thermal velocity, otherwise run-away 
charge carriers would be produced and the conductivity would be much
larger.

The induced electric current is ${\bf J}=en_e {\bf V}$, where 
$n_e$ is the number density of charge carriers. 
Comparing this with the definition of current 
conductivity $\kappa = J/E$ we find:
\be
\kappa = {3\over2 \alpha}\,{n_e \over n_\gamma}{m_e^{2} \over T}
\label{kappa}
\ee
The conductivity is very high, so the generation of magnetic field by the
source currents, created by the cosmological inhomogeneities, 
is governed by the well know equation of magnetic hydrodynamics:
\be
\dt {\bf B} = {\bf \nabla} \times \left( {\bf v} \times {\bf B} \right) + 
{1\over \kappa}\, {\bf \nabla} \times {\bf J}
\label{dt-B}
\ee
The electric current $J$ induced by the
relaxation of the density inhomogeneities
would contain two components: electronic and protonic. However
the first one is surely dominant because it is much easier to drift 
electrons than heavier protons. This is why a non-zero
current can be induced. Of course motion of electrons should 
not produce any excess of electric charge but 
this can be realized because the current 
is created in the dominant homogeneous part of the charge particle 
distribution.

The solution of eq. (\ref{dt-B}) can be roughly written as
\be
B\sim \int_0^t dt_1 \left({2\pi\,J \over \lambda\,\kappa}\right)\,
e^{ 2\pi\,v\,t_1 /\lambda }
\label{b1}
\ee 
The exponential factor under the integral, 
which presents pregalactic dynamo effect 
is normally rather weak, the exponent is about 
\be
{2\pi v t\over\lambda} 
\approx 500\,T_{\rm eV}\, 
\left({\delta T \over T}\right)_\lambda 
\label{expo}
\ee
where eqs. (\ref{v0}) and (\ref{l/t}) have been used and the density 
fluctuations were taken as 
$\delta_\lambda = 4 \left({\delta T / T}\right)_\lambda$. With 
$\left({\delta T / T}\right)_\lambda \sim 3\cdot 10^{-5}$ and 
$T\sim $ eV the exponent is about $0.015 \, T_{\rm eV} <1$   
and pregalactic dynamo is not important. It may be significant if the 
density perturbations at relatively small scales are  much larger than 
their accepted canonical values. Exponential enhancement can be large
at higher temperatures, $T>10^3$ eV, which correspond to the present
day scales below kpc. Equipartition between magnetic field 
and CMBR can be expected on such scales and after 'Brownian'' line
reconnection and relatively mild galactic dynamo the observed 
galactic magnetic fields may be created.

An estimate of magnetic field without pregalactic dynamo enhancement
can be easily done if the helical source current is known,
${\bf \nabla \times J} = e n_e {\bf \Omega}$. With 
${\bf \Omega}$ given by
eq.~(\ref{omega-fin}) and $B$ by eq.~(\ref{b1}) we obtain:
\be
{B_0\over T^2}= 0.24\cdot 10^3\left( 4\pi\alpha\right)^{3/2}\,
\left({t\over \lambda}\right)\,
\left({l_\gamma \over \lambda}\right)^3\, 
\left({ T \over m_e }\right)^2 \approx 
10^{-8} T_{\rm eV}^3
\label{B/T2}
\ee
where we took the wavelength equal to the photon mean free path,
$\lambda = l_\gamma$.

If we take into account that linear compression of pregalactic medium 
in the process of galaxy formation is approximately $r\sim 10^2$, 
the seed field in a galaxy after its formation would be 
$r^2 B_0$, i.e. 4 orders 
of magnitude larger than that given by eq.~(\ref{B/T2}) and, for
$T= 1$ eV, a relatively
mild galactic dynamo, about $10^{4}$, is necessary to obtain 
the observed galactic 
magnetic field of a few micro-Gauss at the scale 
$l_B \sim (100/r)$ kpc $ =1$ kpc. 
The seed magnetic fields formed earlier (at 
higher $T$) would have larger magnitude ($\sim T^3$) but their 
characteristic scale would be smaller by factor $1/T^2$. Chaotic line 
reconnection could create magnetic field at larger, galactic scale 
$l_{gal}$, but the 
magnitude of this field would be suppressed by Brownian motion law -
it would drop by the factor $(l_B/l_{gal})^{3/2}$. It is interesting
that according to our results all scales give comparable contributions
at $l_{gal}$. This effect may lead to an enhancement of the field but 
it is difficult to evaluate the latter.
Let us also note that magnetic fields generated by the discussed
mechanism at the cluster scale, 10 Mpc, should be not larger than
$10^{-8}$ $\mu$G if no additional amplification took place.

\section{Discussion and conclusion \label{s-disc}}

We have shown in this work that photon diffusion produced by temperature 
fluctuations of CMBR could create vortical flows in the primeval plasma
in the second order, i.e. proportional to $(\delta T/T)^2$. Because of
different mobility of electrons and protons (or helium ions) in the plasma
the helical macroscopic motion of the latter would create helical electric
currents which would in turn generate magnetic fields. The amplitude of 
such fields generated at the moment when plasma temperature was equal to
$T$ is given by eq. (\ref{B/T2}). The characteristic scale of the field
should be equal to the photon mean free path in the plasma at temperature
$T$ which at the present day is about 100 kpc $({\rm eV}/T)^2$. This scale
is quite close to the galactic size. Adiabatic compression in the process
of galaxy formation by the factor $r\approx 100$ would lead to an enhancement
of magnetic field by $r^2 = 10^4$. Thus to meet the observational data
this seed field should be amplified by the galactic dynamo only by 4-5 orders
of magnitude. Even if the magnitude of the seed field presented above is
somewhat overestimated there is a huge reserve in the 
amplification by the galactic dynamo which could be as large as 15 orders 
of magnitude~\cite{dynamo}. 

A nice feature of this mechanism is that it does not demand any new
physics for the realization. Of course, density/temperature 
perturbations at large scales were, most probably, created during
inflationary epoch and in this sense new physics is invoked but 
independently of the mechanism we know from observations that 
$\delta T/T \neq 0$ and, based on these data, we may estimate the seed 
magnetic fields using good old physics.

It is interesting that, according to the existing indications to 
intergalactic magnetic fields, their strength is 3-4 orders of magnitude
weaker than the strength of galactic fields. This fact (if it is a fact)
hints that intergalactic and galactic magnetic fields might have common 
origin, but galactic fields are larger due to mentioned above adiabatic 
compression which enhanced the field by 4 orders of magnitude. If it is
so, the mechanism discussed in this work may be irrelevant because
some galactic dynamo is needed to amplify the galactic seed field up to 
the observed magnitude. On the other hand, such dynamo seemingly does not 
operate on intergalactic scales.

Larger density perturbations would be helpful for generation of larger 
magnetic field for which dynamo might be unnecessary. Though much bigger
$\delta T$ is not formally excluded at the scale about 100 kpc, but 
to have them at the level $(\delta T/T)^2\sim 10^{-4}$ seems to be too
much. A natural idea is to turn to a later stage, to onset of structure 
formation when $\delta \rho/\rho$ becomes larger than $10^{-2}$. With such
density perturbations strong enough magnetic fields may be generated 
without dynamo amplification. However after recombination the number 
density of charge carriers drops roughly by 5 orders of magnitude.
Correspondingly $l_\gamma$ rises by the same amount and the strength of
the seed field would be 5 orders of magnitude smaller if density 
perturbations and the temperature of formation remained the same. However
both became very much different. Density perturbations rose as
scale-factor,
$(\delta\rho /\rho)^2\sim (T_{eq}/T)^2$, where $T_{eq}\sim 1$ eV is the 
temperature when radiation domination changed into matter domination and 
density perturbations started to rise. Since, $B/T^2\sim T^3$, according
to eq.~(\ref{B/T2}), the net effect of going to smaller $T$ is a decrease
of $B/T^2$ which would be difficult to cure even by later reionization. 
Still, as argued in ref.~\cite{langer03}, magnetic field generation,
driven by anisotropic and inhomogeneous radiation pressure (and in this
sense similar to our mechanism) at the epoch of reionization, could end 
up with the field of about $8\cdot 10^{-6} \mu{\rm G}$. This result is 
8 orders of magnitude larger than that found in the earlier 
papers~\cite{reion} and quite close to ours (\ref{B/T2}), though these 
two mechanisms operated during very different cosmological epochs and 
were based on different physical phenomena. 

After this paper was sent to astro-ph we became aware of the
work~\cite{hogan} "Generation of Cosmic Magnetic Fields at Recombination"
where a much weaker effect was found. We think that the difference can be
attributed to the following effects. We considered earlier period when the
photon mean free path was much smaller than the horizon. It gives a factor
about $10^3$ in fluid velocity, eq.(\ref{v0}). Moreover, since in our
case the electrons are tightly bound to photons the electron-photon
fluid moves as a whole (while protons and ions are at rest) and the
electric current induced by macroscopic motion/oscillations of plasma
is noticeably larger.

\vspace{6mm} 
{\large \bf Acknowledgments} 

We thank C. Hogan and P. Naselsky for critical comments.
This work was partially supported by the MIUR research grant 
"Astroparticle Physics".

\end{document}